\documentclass[letterpaper,12pt]{amsart}

	\usepackage{amsmath, amsfonts, amsthm, mathrsfs, amssymb, graphicx, subfigure}

	\usepackage{amsaddr}
	
	\theoremstyle{plain}

	\title{Microscopic Origins of Macroscopic Behavior}
	\author{Joel L. Lebowitz}
	\address{Departments of Mathematics and Physics\\
	Rutgers University}
	\email{lebowitz@math.rutgers.edu}
	\date{}
\begin{document}

	%\addtocounter{page}{-1}

	%\begin{frame}[plain]
	
	%\titlepage
	
	\maketitle
	
	%\begin{center}
	%(Much of the work described here was done jointly with S. Goldstein, R. Tumulka, D. Huse and N. Zanghi)
	%\end{center}
	
	%\end{frame}

	%\begin{frame}\frametitle{}
	
	This article is mostly based on a talk I gave at the March 2021 meeting (virtual) of the American Physical Society on the occasion of receiving the Dannie Heineman prize for Mathematical Physics from the American Institute of Physics and the American Physical Society. I am greatly indebted to many colleagues for the results leading to this award. To name them all would take up all the space allotted to this article. (I have had more than 200 collaborators so far), I will therefore mention just a few: Michael Aizenman, Bernard Derrida, Shelly Goldstein, Elliott Lieb, Oliver Penrose, Errico Presutti, Gene Speer and Herbert Spohn. I am grateful to all of my collaborators, listed and unlisted.	
	%\ 
	
	I would also like to acknowledge here long time support form the AFOSR and the NSF.
	
	\ 
	
	\noindent\textbf{\underline{Introduction}}
	
	Let me begin by quoting Freeman Dyson, an earlier recipient of this prize, about his definition of mathematical physics:
	\begin{quote}
	``Mathematical physics is the discipline of people who try to reach a deep understanding of physical phenomena by following the rigorous style and method of mathematics.''
	\end{quote}
	\begin{flushright}
	--- Freeman Dyson, \textit{From Eros to Gaia}, p 164-165
	\end{flushright}
	Freeman was a good friend and this talk is dedicated to his memory, as well as to the memory of my teachers Melba Phillips and Peter Bergmann.
	
	My own area of mathematical physics is statistical mechanics, which is concerned with the microscopic origin of macroscopic behavior. Since our mathematical abilities for dealing with strongly interacting many particle systems are quite limited it is fortunate that many striking features of macroscopic systems can be obtained from simplified microscopic models.
	
	We therefore often take as our lowest level starting point an idealized description of atoms. As put by Feynman \cite{Fe63}:
	\begin{quote}
	``If in some cataclysm all of scientific knowledge was to be destroyed, \ldots, what statement would contain the most information in the fewest words? I believe it is \ldots that all things are made of atoms --- little particles that move around in perpetual motion, attracting each other when they are a little distance apart, but repelling upon being squeezed into one another.''
	\end{quote}
	
	Unfortunately even such idealized systems are too difficult to deal with in any detail. In fact many details would just be confusing. I will therefore focus on describing, in a qualitative way, the microscopic origin of those behaviors which are (almost) always observed in isolated macroscopic systems both in equilibrium and out of it. I will relate this to the fact that this behavior is \textit{typical} for systems represented by the usual Gibbs measures or those derived from them. These take small phase space volume to indicate small probability. I will not try to justify this here.
	
	This means that for equilibrium macroscopic systems these behaviors occur for an overwhelming majority of the microstates in the micro-canonical ensemble, i.e. they are typical. In fact, the fraction of systems with noticeable macroscopic deviations from the average behavior, computed in such an ensemble, is exponentially small in the number of degrees of freedom of the system: the functions on the phase space which correspond to such typical behavior will be described later.
	
	An analogous statement holds for the (exponentially small) subsets of the micro-canonical ensemble which describe systems in nonequilibrium macrostates (to be defined below). It thus includes the time asymmetric approach to equilibrium, encoded in the second law and observed in individual macroscopic systems. Once one accepts the applicability of these measures to physical systems the observed behavior does not require explanations based on ergodicity, time averaging, or subjective information theory.
	
	This property of typicality of behavior predicted by the measures used to represent macroscopic systems is true both classically and quantum mechanically. It explains why these ensembles can be used to predict the observed behavior of individual macroscopic systems \textit{and} not just some average behavior.
	
	I will begin with classical systems where the situation is easier to visualize.
	
	\ 
	
	\noindent\textbf{\underline{Classical Systems}}
	
	In classical mechanics, the microstate of a system of \( N\) particles confined to a region \( V\) in \(\mathbb R^ d\) is a point \( X\) in the \(2 d N\)-dimensional phase space,
	\begin{align}
	X=(\vec r_1,\vec p_1,\ldots,\vec r_ N,\vec p_ N),\quad \vec r_ i\in V\subset \mathbb R^ d,\quad\vec p_i\in\mathbb R ^ d
	\end{align}
	Its time evolution is given by a Hamiltonian \( H( X)\) which conserves energy, so \(X(t)\) will be confined to \(\Gamma_E\), a thin shell surrounding the energy surface \(H(X)=E\). A macroscopic system is one with ``very large'' \( N\), say \( N\gtrsim10^{20}\).
	
	\ 
	
	\noindent\textbf{\underline{Macrostates}}
	
	To describe the macroscopic state of such a system of \( N\) particles in a box \( V\), we make use of a much cruder description than that provided by the microstate \( X\). We shall denote by \( M\) such a macroscopic description: \( M( X)\) is the macrostate of the system in the microstate \( X\). As an example we may divide \( V\) into \( K\) cells, where \( K\) is large but still \( K\ll N\), and specify the number of particles, the momentum and the amount of energy in each cell, with some tolerance. Clearly there are many \( X\)'s (in fact a continuum) which correspond to the same \( M\). Let \(\Gamma_ M\) be the region in \(\Gamma_ E\) consisting of all microstates \( X\) corresponding to a given macrostate \( M\) and denote by \(|\Gamma_ M|\) its Liouville volume.
	
\begin{figure}
	\centering
	\includegraphics[width=3in]{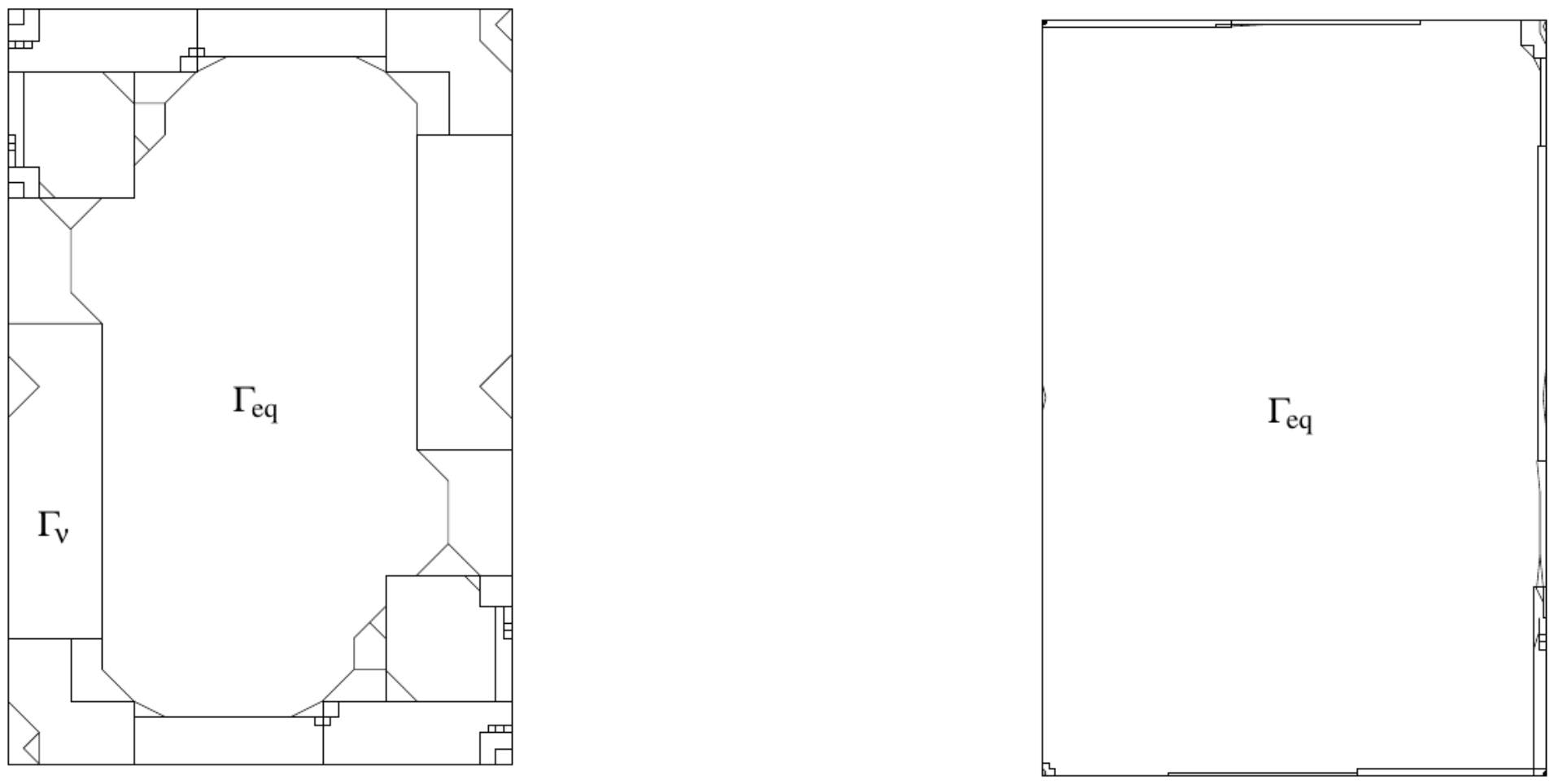}
	\caption{Schematic Picture of the decomposition of \(\Gamma_E\).}
	\label{fig1}
	\end{figure}

	It can be proven \cite{La73} that, generally, that in every \(\Gamma_ E\) of a macroscopic system there is one region \(\Gamma_ M\) which has most of the volume of \(\Gamma_E\). This is called the equilibrium macrostate \( M_{\mathrm{eq}}\),
	\begin{align}
	\frac{|\Gamma_{ M_{\mathrm{eq}}}|}{|\Gamma_E|}=1-\varepsilon
	\end{align}
	with \(\varepsilon\ll1\). When \( M( X)\) specifies a nonequilibrium state, \(|\Gamma_{ M}|\) is much smaller. Thus for a gas consisting of \( N\) particles in a volume \( V\) the ratio of \(|\Gamma_ M|\), the volume of a macrostate \( M\) in which all the particles are in the left half of the box, and \(|\Gamma_{ M_{\mathrm{eq}}}|\), the volume of the macrostate \( M_{\mathrm{eq}}\) in which there are \((\frac{1}{2}\pm10^{-10}) N\) particles in the left half of the box, is of order \(2^{-N}\); see Figure \ref{fig1}.
	The second picture is slightly more faithful. Neither shows the topology or differences in relative sizes of the different \(\Gamma_ M\)'s. In general, the closer \( M\) is to \( M_{\mathrm{eq}}\) the larger \(\Gamma_ M\).
	
	A system is then in macroscopic thermal equilibrium iff \( X\in\Gamma_{ M_{\mathrm{eq}}}\).
	
	Relevant properties of macroscopic systems depend only on sums over the entire system of functions which depend only on the coordinates and momenta of a few particles. The values of the sums, suitably scaled with \(N\), are approximately the same for almost all \( X\in\Gamma_{ M_{\mathrm{eq}}}\), hence they are typical and observed for (almost) all systems in equilibrium. In fact since \(|\Gamma_{ M_{\mathrm{eq}}}|\sim|\Gamma_ E|\) they are also typical of \( X\in\Gamma_ E\).
	
	This justifies the use of the microcanonical ensemble to compute relevant properties of an equilibrium system: independent of whether or not the dynamics is ergodic in a mathematical sense.
	
	\ 
	
	\noindent\textbf{\underline{Nonequilibrium States}}
	
	Thus, most microstates in \(\Gamma_ E\) of a macroscopic system correspond to the system being in equilibrium. A similar statement is true for most wave functions, in fact also for most energy eigenstates, in \(\mathscr H_ E\), the energy shell of the Hilbert space: see later. Fortunately there are also microstates which correspond to macroscopic systems which are out of equilibrium (or we would not be here).
	
	Given an \( X( t_0)\) in such a \(\Gamma_ M\), \( M\ne M_{\mathrm{eq}}\), at an ``initial'' time \( t_0\), we would like to know how the properties of a macroscopic system, isolated for \( t> t_0\), change with time.
	
	\ 
	
	\noindent\textbf{\underline{Approach to Equilibrium}}
	
	Boltzmann (also Maxwell, Kelvin, \ldots) argued that given the disparity in the sizes of the \(\Gamma_ M\) corresponding to the various macrostates, the evolution of a ``typical'' microstate \( X\), specified to be at \( t= t_0\) in the phase space region \(\Gamma_ M\), will be such that \(|\Gamma_{ M( X( t))}|\) will not decrease (on a macroscopic scale) for \( t> t_0\). In fact for any \(\Gamma_ M\) the relative volume of the set of microstates \( X\in\Gamma_ M\) for which this is false over some time period \(\tau\) during which the macrostate undergoes a macroscopically noticeable change, but not longer than the age of the universe, goes to zero exponentially in the number of atoms in the system. This explains and describes the evolution towards equillibrium of macroscopic systems which start in the macrostate \(\Gamma_ M\), \( M\ne M_{\mathrm{eq}}\), and are kept (effectively) isolated afterwards.
	
	\ 
	
	\noindent\textbf{\underline{Boltzmann's Entropy}}
	
	To make a connection with the Second Law, Boltzmann defined the (Boltzmann) entropy of a macroscopic system in a microstate \( X\) as
	\begin{align}\label{ent}
	S_{\mathrm B}( X)=\log|\Gamma_{ M( X)}|= S_{\mathrm B}( M).
	\end{align}
	Boltzmann then showed that the Clausius thermodynamic entropy of a gas in equilibrium is equal to \(\log|\Gamma_ E|\sim\log|\Gamma_{ M_{\mathrm{eq}}}|\).
	
	The above heuristic argument, based on relative phase space volume, is the correct explanation for the behavior typically observed in actual macroscopic systems. It is, however, very far from a mathematical theorem and contains no quantitative information about time scales. A desirable result would be the rigorous derivation from the microscopic dynamics of the kinetic and hydrodynamic equations commonly used to describe the time asymmetric, entropy increasing, observed behavior of macroscopic systems out of equilibrium. This has been achieved so far only for the Boltzmann equation for dilute gases. This was done rigorously (in appropriate limits) by Oscar Lanford in 1975. (I will not discuss derivations which include some external randomness in the dynamics by the Varadhan school. I will also not discuss the derivation of a diffusion equation for non-interacting particles moving among Sinai billiards. Those cases show what we could do if only our mathematics was better.)
	
	\ 
	
	\noindent\textbf{\underline{The Boltzmann Equation for Dilute Gases}}
	
	Following Boltzmann, we refine the description of a macrostate \( M\) by noting that the microstate \( X=\{\mathbf r_ i,\mathbf v_ i\}\), \( i=1,\ldots, N\), can be considered as a set of \( N\) points in six dimensional one particle space. We then divide up this one particle space into \( J\) cells \(\Delta_\alpha\), centered on \((\mathbf r_\alpha,\mathbf v_\alpha)\), of volume \(|\Delta_\alpha|\). A macrostate \( M_f(X)\) can then be specified by a distribution \(f(\textbf{x},\textbf{v})\) in the one-particle space such that the fraction of particles in each \(\Delta_\alpha\), is given by
	\begin{align}
	N_\alpha/N\cong\int_{\Delta_\alpha}\mathrm d\mathbf x\mathrm d\mathbf v\ f(\mathbf x,\mathbf v).
	\end{align}
	
	Boltzmann then used his deep physical intuition (and hints from Maxwell) to derive his eponymous equation for the time evolution of the macrostate \(M_f(X(t))\) given by \(f(\mathbf x,\mathbf v, t)\). I will not reproduce this equation here as this can be found in all textbooks on kinetic theory, c.f. \cite{Ce94}. The reasoning behind it is carefully explained in Lanford's beautiful non-technical article \cite{La76}, one of the best mathematical-physics articles I have ever read. I strongly recommend it.
	
	I will now give a bird's eye view of what I think is the essence of that article:
	
	Let \(f_0(\mathbf x, \mathbf v),\mathbf x\in V\subset\mathbb R^3,\mathbf v\in\mathbb R^3,\) be a smooth function of integral one. Then consider a gas consisting of \(N\) spheres of diameter \(d\) in \(V\). Keeping \(V\) fixed consider now a sequence of states with different particle numbers, \(N\to\infty\), \(d\to0\), such that \(Nd^2\to b>0\), while \(Nd^3\to0\). This is called the Boltzmann-Grad (BG) limit. Consider now all the phase points \(X_N\) of this gas such that  the fraction of particles in \(\Delta_\alpha\), satisfies \(N_\alpha/N\cong\int_{\Delta_\alpha}f_0(\mathbf x,\mathbf v)\mathrm d\mathbf x\mathrm d\mathbf v\) with,
	\begin{align}
	\lim_{\mathrm{BG}}N_\alpha/N=\int_{\Delta_\alpha}f_0(\mathbf x,\mathbf v)\mathrm d\mathbf x\mathrm d\mathbf v
	\end{align}
	The system with \(N\) particles evolves according to Hamiltonian dynamics for elastic collisions going from \(X_N\) to \(X_N(t)\), \(t>0\). Lanford's theorem then says:
	
	There exists a \(\tau>0\) such that for \(t<\tau\) the \(N_\alpha(t)\) corresponding to asymptotically almost all such \(X_N(t)\) satisfy
	\begin{align}\label{lim}
	\lim_{BG}\frac{N_\alpha( t)}{N}=\int_{\Delta_\alpha}f(\mathbf x,\mathbf v,t)\mathrm d\mathbf x\mathrm d\mathbf v
	\end{align}
	where \(f(\mathbf x,\mathbf v,t)\) evolves according to the Boltzmann equation with initial condition \(f_0\). Here again almost all is with respect to the relative phase space volume. \eqref{lim} holds for all reasonable sets of \(\Delta_\alpha\)'s.
	
	The time \(\tau\) for which Lanford's theorem holds is about one fifth of the mean free time between collisions, but that is a purely technical problem. This time is long enough for the Boltzmann entropy per particle of the macrostate \(M_f\) to increase by a finite amount.
	
	The Boltzmann entropy of the macrostate \(M_f\), associated with the distribution \(f\) is defined as in \eqref{ent},
	\begin{align}
	S_{\mathrm B}(f)=S_{\mathrm B}(M_f)=\log|\Gamma_{M_f}|
	\end{align}
	where \(|\Gamma_{M_f}|\) is the phase space volume corresponding to \(M_f\). \(S_{\mathrm B}(f)\) was actually computed by Boltzmann. He showed that, up to constants, this is given for a dilute gas, by
	\begin{align}
	\frac{1}{N}S_{\mathrm{B}}( f)=-\int_V\mathrm d\mathbf x\int_{\mathbb R^3}\mathrm d\mathbf v\  f(\mathbf x,\mathbf v)\log f(\mathbf x,\mathbf v)
	\end{align}
	This agrees with the Gibbs-Shannon entropy per particle for a system in a product measure, with each particle having distribution \(f(\mathbf x,\mathbf v)\), but is conceptually not the same at all (see below).
	
	The maximum of \( S_{\mathrm{B}}( f)\) over all \( f\) with a given energy, which is here just the kinetic energy, is given by the Maxwell distribution
	\begin{align}
	f_{\mathrm{eq}}=\frac{ N}{|V|}(2\pi k T/ m)^{-3/2}\exp[-m\mathbf v^2/2 k T]
	\end{align}
	where \( k T=2/3( E/ N)\).
	
	In this case
	\begin{align}
	\frac{1}{N}S_{\mathrm{B}}( f_{\mathrm{eq}}) = \frac{3}{2}\log T-\log\frac{ N}{| V|}+\mathrm{Const.}
	\end{align}
	the same as the Clausius entropy for a dilute gas.
	
	When \( f\ne f_{\mathrm{eq}}\) then \( f\) and consequently \( S_{\mathrm{B}}( f)\) will change in time.
	
	The second law, now says that for \textit{typical} \( X\), \( f_{ X_ t}(\mathbf x,\mathbf v)= f(\mathbf x,\mathbf v,t)=f_t\) has to be such that \( S_{\mathrm{B}}( f_ t)\ge S_{\mathrm{B}}( f_{t'})\), for \( t\ge t'\).
	
	This is exactly what happens for a dilute gas described by the Boltzmann equation.
	\begin{align}
	\frac{\operatorname d}{\operatorname d t}S_{\mathrm{B}}( f_ t)\ge0,\quad \text{Boltzmann's } \mathcal H\text{-theorem}
	\end{align}
	As put by Boltzmann \cite{Bo98}:
	\begin{quote}
	``In one respect we have even generalized the entropy principle here, in that we have been able to define the entropy in a gas that is not in a stationary state.''
	\end{quote}
	
	\ 
	
	\noindent\textbf{\underline{More General Hydrodynamic Equations}}
	
	Suppose, more generally, that the time evolution of the macrostate \( M\), given by \( M( X( t))= M_ t\), effectively satisfies an autonomous deterministic time asymmetric equation, such as the Navier-Stokes equation or the heat equation or the Boltzmann equation just discussed.
	
	Such an equation means that if \( M_{ t_1}\to M_{ t_2}\), for \( t_2> t_1\), and \( M_{ t_2}\to M_{ t_3}\), for \( t_3> t_2\), then the microscopic dynamics \( T_ t\) carries \(\Gamma_{ M_{ t_1}}\) inside \(\Gamma_{ M_{ t_2}}\), i.e.\ \( T_{ t_2- t_1}\Gamma_{ M_{ t_1}}\subset\Gamma_{ M_{ t_2}}\) and \( T_{ t_3- t_2}\Gamma_{ M_{ t_2}}\subset\Gamma_{ M_{ t_3}}\), with \textit{negligible error}. Put otherwise a typical phase point in \(\Gamma_{ M_{t_1}}=\Gamma_{ M_1}\) will go to \(\Gamma_{ M_2}\) and then to \(\Gamma_{ M_3}\), i.e.\ \( T_{ t_3- t_1}\Gamma_{M_{1}}\subset\Gamma_{M_3}\).
	
	\begin{figure}
	\centering
	\includegraphics[width=2in]{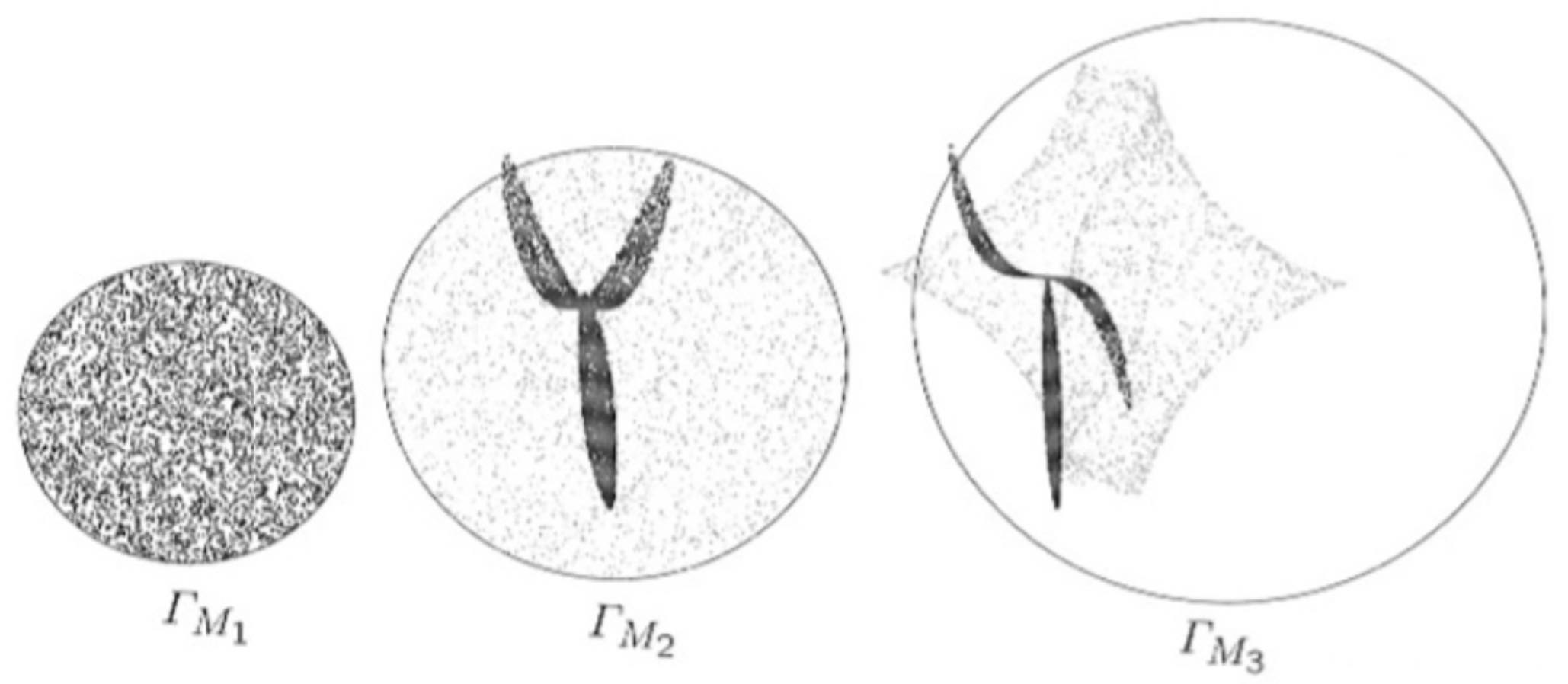}
	\caption{Time evolution of a macrostate}
	\label{fig2}
	\end{figure}
	
	The fact that phase space volume is conserved by the Hamiltonian time evolution implies that \(|\Gamma_{ M_{1}}|\le|\Gamma_{ M_{2}}|\le|\Gamma_{ M_{3}}|\), and thus that \( S_{\mathrm{B}}( M_{3})\ge S_{\mathrm{B}}( M_{2})\ge S_{\mathrm{B}}( M_{1})\). A deterministic macroscopic equations, for increasing time, then has to satisfy the inequality \linebreak \(\frac{\mathrm d}{\mathrm d t}S_{\mathrm B}(M_t)\ge0\), \cite{Go04,Pe70}.
	
	A crucial point here is that the phase points in the region in \(\Gamma_2\) coming from \(\Gamma_1\) behave, forward in time, as microstates typical of \(\Gamma_2\). They are, however, very atypical backwards in time: if we reverse all the velocities in \(\Gamma_2\), then at a later time, \(t'=t_2+(t_2-t_1)\) all of the points initially in \(\Gamma_{M_1}\) will again be in \(\Gamma_{M_1}\) (with their velocities reversed) a smaller region than \(\Gamma_{M_2}\). (The same is true for any sequence of positive times.)
	
	The reason for this asymmetry in typical behaviors is due to initial conditions. That is, when nature or the experimentalist who is part of nature, starts out with a nonequilibrium system in an initial state \(X\in\Gamma_M\) we can assume that \(X\) is typical of \(\Gamma_{M}\), and continues to be so in the forward time direction.
	
	But how did all this get started? In the Lanford derivation reversing the velocities at some \(t<\tau\) violates the assumptions on the initial conditions required for the derivation of the Boltzmann equation. But what about real life situations? Somewhat surprisingly, if one thinks about it, one has to go back to the very beginning of the world we live in. This was already fully understood by Boltzmann and others as the quotes below show.
	
	\ 
	
	\noindent\textbf{\underline{Initial Conditions}}
	
	\begin{quote}
	``From the fact that the differential equations of mechanics are left unchanged by reversing the sign of time without changing anything else, Herr Ostwald concludes that the mechanical view of the world cannot explain why natural processes always run preferentially in a definite direction. \textbf{But such a view appears to me to overlook that mechanical events are determined not only by differential equations, but also by initial conditions.} In direct contrast to Herr Ostwald I have called it one of the most brilliant confirmations of the mechanical view of Nature that it provides an extraordinarily good picture of the dissipation of energy, \textbf{as long as one assumes that the world began in an initial state satisfying certain conditions. I have called this state an improbable state.}''
	\end{quote}
	\begin{flushright}
	--- L. Boltzmann \cite{Bo97}
	\end{flushright}
	
	\begin{quote}
	``It is necessary to add to the physical laws the hypothesis that, in the past the universe was more ordered in the technical sense, [i.e.\ low \(S_\mathrm B\)] than it is today \ldots to make an understanding of irreversibility.''
	\end{quote}
	\begin{flushright}
	--- R.P. Feynman \cite{Fe67}
	\end{flushright}
	
	\begin{figure}[h]
	\centering
	\includegraphics[width=4in]{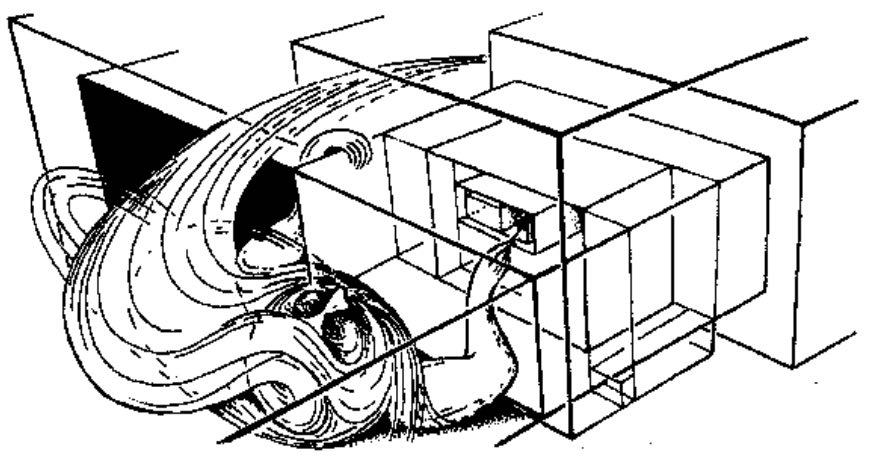}
	\caption{``Creation of the universe: a fanciful description! The Creator's pin has to find a tiny box, just 1 part in \(10^{10^{123}}\) of the entire phase-space volume, in order to create a universe with as special a Big Bang as we actually find.'' from R. Penrose, \textit{The Emperor's New Mind} \cite{Pe90}}
	\label{fig3}
	\end{figure}
	
	The ``tiny box'' in Fig. 3 is a macrostate with low \(S_\mathrm B\). N.B. It is not necessary to select a particular microstate. Almost all microstates in a low-entropy macrostate will behave in a similar way.
	
	It may be relevant to mention here a question I was asked during a talk I gave on the subject: Q: What does the initial state of the universe have to do with the fact that when I put my sugar cube in my tea it dissolves irreversibly? A: Nothing directly but the fact that you, the sugar cube and the tea are all here is a consequence of the initial low entropy state of the universe.

	\ 
	
	\noindent\textbf{\underline{Boltzmann vs. Gibbs Entropies}}
	
	Given an ensemble (probability) density \(\mu(X)\), the Gibbs-Shannon entropy is given by
	\begin{align}
	 S_{\mathrm G}\equiv- k\int_\Gamma\mu\log\mu\ \mathrm d X.
	\end{align}
	
	Clearly if \(\mu=\tilde\mu_ M\), where
	\begin{align}
	\tilde\mu_ M=\begin{cases}
	\ |\Gamma_ M|^{-1},&\text{ if } X\in\Gamma_ M;\\
	\quad0,&\text{ otherwise }
	\end{cases}
	\end{align}
	then
	\begin{align}
	 S_{\mathrm G}(\tilde\mu_ M)= k\log|\Gamma_ M|= S_{\mathrm B}( M).
	\end{align}
	This is essentially the case for the microcanonical ensemble and thus the Gibbs and Boltzmann entropies are equal for equilibrium systems.
	
	However \( S_{\mathrm G}(\mu)\) does not change in time for isolated systems and therefore is ``useless'' for such systems not in equilibrium, while \( S_{\mathrm B}( M( X))\) captures the essence of typical macroscopic behavior.
	\begin{figure}
	\centering
	\includegraphics[width=3.7in]{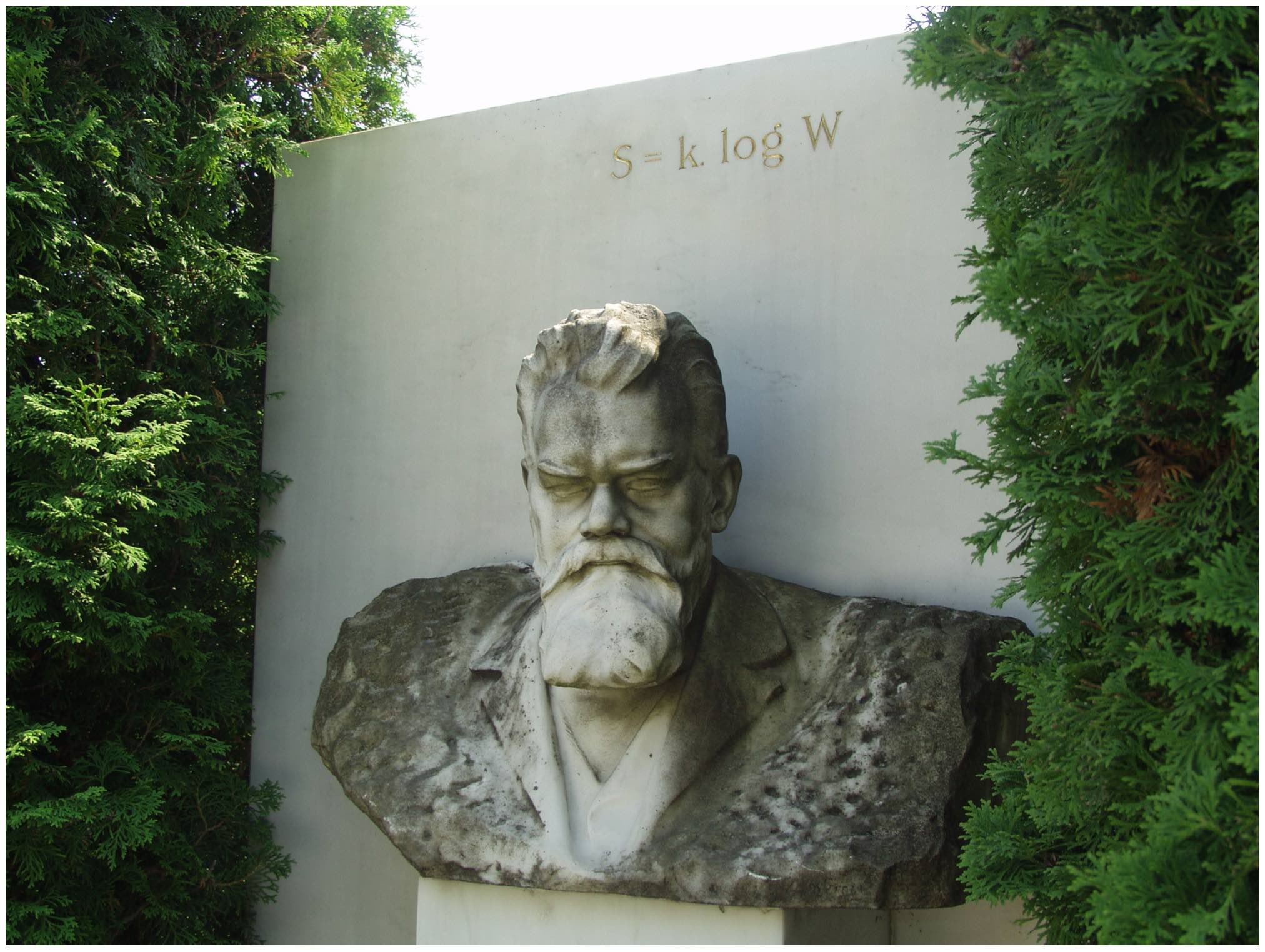}
	\caption{Boltzmann's grave in Zentralfriedhof, Vienna, with bust and entropy formula}
	\label{fig4}
	\end{figure}
	
	\ 
	
	\noindent\textbf{\underline{Quantum Systems}}
	
	Let me turn now to quantum systems.
	
	I will take the microstate of a system to be its wave function \(\psi\in\mathscr H_ E\), where \(\mathscr H_ E\) is a shell in Hilbert space of thickness \(\Delta E\), i.e. \(\psi\) is a linear combination of energy eigenfunctions in the range \((E,E+\Delta E)\), \(\Delta E\ll E\) but very large compared to the spacing between levels.
	
	This is not perfect (problems with Schr\"odinger's Cat) but it will have to do for the present (see below).
	
	The question then is which \(\psi\) correspond to the system being in macroscopic thermal equilibrium (MATE), i.e.\ what is the analog of a classical microstate \( X\) lying in \(\Gamma_{ M_{\mathrm{eq}}}\).
	
	Following von Neumann (Chapter 5 of \cite{Ne55}), we take the macro-observables \(M\) to commute with each other. We assume that this can be achieved by suitably ``rounding off'', i.e.\ coarse-graining, the operators representing the macro-observables.
	
	The coarse-grained energy operator commutes with the other coarse-grained macro-observables. Thus all \(M\)'s can be regarded as operators on \(\mathscr H_ E\). Their joint spectral decomposition defines an orthogonal decomposition
	\begin{align}
	\mathscr H_ E=\bigoplus_\nu\mathscr H_\nu,
	\end{align}
	
	The subspaces \(\mathscr H_\nu\) (``macro spaces''), the joint eigenspaces of the macro-observables, correspond to the different macro states. This corresponds to the division of the classical energy shell \(\Gamma_E\) into disjoint regions \(\Gamma_ M\).
	
	A system is in a macrostate \(M_\nu\) if its wave function \(\psi\) is ``close'' to \(\mathscr H_\nu\), i.e.\ \(\langle\psi| P_\nu|\psi\rangle\ge1-\delta\), \(\delta\ll1\), with \(P_\nu\) being the projection to \(\mathscr H_\nu\). As noted earlier due to the Schr\"odinger's Cat problem there will be \(\psi\) which are a superposition of \(\psi\)'s in different macrostates. To remedy this one has to go beyond the Copenhagen interpretation of the wave function being a complete description, c.f. \cite{Be87,Go98}. For the present let me say that I would interpret such \(\psi\)'s as giving probabilities of being in different macrostates.
	
	The ``volume'' of each macro space \(\mathscr H_\nu\) is its dimension \(d_\nu\).
	
	As in the classical case, it is generally true that one of the \(\mathscr H_\nu\), denoted \(\mathscr H_{\mathrm{eq}}\), has most of the dimensions of \(\mathscr H_ E\), i.e.,
	\begin{align}\label{argument}
	\frac{\dim\mathscr H_{\mathrm{eq}}}{\dim\mathscr H_ E}=1-\varepsilon
	\end{align}
	with \(\varepsilon\ll1\).
	
	A macroscopic system is in MATE if
	\begin{align}
	\langle\psi| P_{\nu_\mathrm{eq}}|\psi\rangle\ge1-\delta.
	\end{align}
	
	The Boltzmann entropy \( S_\mathrm{B}(\psi)\) of a system in a macrostate \(M\) is then given by the log of the dimension of the macro space \(\mathscr H_\nu\); \(S_\mathrm B(M_\nu)=\log d_\nu\): \(d_\nu\) is the quantum analogue of \(|\Gamma_{M_\nu}|\).
	
	Most everything we said about isolated classical macroscopic systems then also holds for quantum systems. In particular for \(\varepsilon\ll\delta\), most pure states in \(\mathscr H_E\) are in MATE.
	
	When I say ``for most \(\psi\)'', I mean that \(\psi\) is typical with respect to a uniform measure on the unit sphere in \(\mathscr H_ E\). This measure on ``wave functions'' was considered already by Schr\"odinger and particularly Felix Bloch \cite{Go06}. It yields the microcanonical measure \(\rho^\mathrm{mc}\) but goes beyond it.
	
	The Boltzmann argument for increase in entropy of isolated macroscopic systems out of equilibrium is then similar to that in the classical case \cite{Gr94}. Unlike classical systems however, where any subsystem of a system in a pure state is also in a pure state, a subsystem \(S\) of a quantum system with a wave function \(\psi\) will be described by a density matrix \(\rho_ S^\psi\).
	
	We can now define \cite{Go17} a system with wave function \(\psi\) to be in microscopic thermal equilibrium (MITE) if, for any not-too-large subsystem \( S\), say subsystems with linear dimension \(\ell<\ell_0\), the reduced density matrix of \( S\) is close to the thermal equilibrium density matrix of \( S\)
	\begin{align}
	\rho^\psi_ S\approx\rho^\mathrm{mc}_ S
	\end{align}
	where
	\begin{align}
	\rho^\psi_ S=\operatorname{tr}_{ S^\mathrm{c}}|\psi\rangle\langle\psi|
	\end{align}
	is the reduced density matrix of \( S\) obtained by tracing out the complement \( S^\mathrm{c}\) of \( S\), and
	\begin{align}
	\rho^\mathrm{mc}_ S=\operatorname{tr}_{ S^\mathrm{c}}\rho^\mathrm{mc}
	\end{align}
	
	\(\rho^\mathrm{mc}\) is the microcanonical density matrix corresponding to a uniform distribution over energy eigenstates in \(\mathscr H_E\). For macroscopic systems \(\rho^\mathrm{mc}_S\) can be replaced by \(\rho^\mathrm{ca}_S\), where \(\rho^\mathrm{ca}\) is the canoncial density matrix.
	
	The distinction between MITE and MATE is particularly relevant for systems with many-body localization (MBL) for which the energy eigenfunctions fail to be in MITE while necessarily most of them, but not all, are in MATE.
	
	The argument for most energy eigenfunctions being in MATE is based on the fact that, calling \(D\) the dimension of \(\mathscr H_E\), we have for energy eigenfunctions \(|n\rangle\)
	\begin{align}
	\frac1 D\sum_{n=1}^D\langle n|P_{\nu_\mathrm{eq}}|n\rangle=\frac1 D\operatorname{tr}(P_{\nu_\mathrm{eq}})=1-\varepsilon
	\end{align}
	Noting that \(\langle n|P_{\nu_\mathrm{eq}}|n\rangle\le1\), the average being close to \(1\) means that most eigenstates are close to \(\mathscr H_{\nu_\mathrm{eq}}\). This is consistent with the Eigenfunction Thermalization Hypothesis (ETH).
	
	In fact for generic macroscopic systems, including those with MBL, most wave functions in an energy shell are in both MATE and MITE.
	
	This follows from the following result.
	
	\ 
	
	\noindent\textbf{\underline{Canonical Typicality}}
	
	Consider an isolated system consisting of two parts. Call them system 1 and 2 or system and reservoir. Then \cite{Go06} we have the following result, see also \cite{Le08}.
	
	Let \( H\) be the Hamiltonian of the whole system and let the number of particles in system 1 and 2 be \( N_1\ll N_2\). Let \(\mathscr H_ E\subset\mathscr H_1\otimes\mathscr H_2\) be an energy shell. Then for most \(\psi\in\mathscr H_ E\) with \(||\psi||=1\),
	\begin{align}
	\operatorname{tr}_2|\psi\rangle\langle\psi|\approx\operatorname{tr}_2\rho^\mathrm{mc},
	\end{align}
	where \(\rho^\mathrm{mc}\) is the microcanonical density matrix of the whole system at energy \( E\), i.e.\ equal weight to all energy eigenstates in \(\mathscr H_ E\).
	
	\ 
	
	The theorem says that most wave functions in the energy shell \(\mathscr H_ E\) are both in MATE and in MITE. In fact for macroscopic systems one can show that MITE implies MATE. The opposite is however not true. This is particularly relevant when one considers energy eigenfunctions \(|n\rangle\). While most energy eigenstates, including those for systems with MBL must, as shown, generally be in MATE, most energy eigenfunctions for systems with MBL are not in MITE.
	
	There is no analog to MITE for a classical system where any subsystem of a composite system in state \( X^{(1,2)}\) is also in a unique state \( X^{(2)}\).
	
	When the interaction between systems 1 and 2 is weak, \( H\approx H_1\otimes I_2+I_1\otimes H_2\), then, as is well known,
	\begin{align}
	\operatorname{tr}_2\rho^{\mathrm{mc}}\approx\frac{1}{ Z} e^{-\beta H_1},
	\end{align}
	for \(\beta=\beta( E)=\mathrm d S_{\mathrm{eq}}( E)/\mathrm d E\).
	
	If MATE-ETH holds strictly, i.e., if \textit{all} energy eigenstates in \(\mathscr H_E\) are in MATE, then every state \(\psi\in\mathscr H_E\) will sooner or later reach MATE and spend most of the time in MATE in the long run. That is because, writing \(\overline{f(t)}=\lim_{T\to\infty}\frac1 T\int_0^T\int f(t)\mathrm dt\) for time averages, \(|n\rangle\) for the energy eigenstate with eigenvalue \(E_n\) and \(\psi_t=e^{-iHt}\psi\),
	\begin{align}
	\overline{\langle\psi_t|P_\mathrm{eq}|\psi_t\rangle}&=\sum_{n,n'}\langle\psi| n\rangle\overline{e^{iE_nt}\langle n|P_\mathrm{eq}|n'\rangle e^{-iE_{n'}t}}\langle n'|\psi\rangle\\
	&=\sum_n|\langle\psi|n\rangle|^2\langle n|P_\mathrm{eq}|n\rangle \ge\sum_n|\langle\psi|n\rangle|^2(1-\delta)\\
	&=1-\delta,
	\end{align}
	provided \(H\) is non-degenerate, i.e., \(E_n\ne E_{n'}\) for \(n=n'\) (using \(\overline{e^{iEt}}=1\) if \(E=0\) and \(=0\) otherwise).
	
	\ 
	
	A similar statement is true when there is degeneracy.
	
	\ 
	
	\noindent\textbf{\underline{Summary of Boltzmann's Ideas (also Maxwell, Kelvin, Feynman)}}
	
	Time-asymmetric behavior as embodied in the second law of thermodynamics is observed in \textit{individual macroscopic} systems. It can be understood as arising naturally from time-symmetric microscopic laws when account is taken of a) the great disparity between microscopic and macroscopic sizes, b) initial conditions, and c) that what we observe is ``typical'' behaviors --- not all imaginable ones. Common alternate explanations, such as those based on equating irreversible macroscopic behavior with ergodic or mixing properties of ensembles (probability distributions) already present for chaotic dynamical systems having only a few degrees of freedom or on the impossibility of having a truly isolated system, are either unnecessary, misguided or misleading.
	
	Let me end this article by quoting Einstein's tribute to Boltzmann.
	\begin{quote}
	``On the basis of kinetic theory of gases Boltzmann had discovered that, aside from a constant factor, entropy is equivalent to the logarithm of the ``probability'' of the state under consideration. Through this insight he recognized the nature of course of events which, in the sense of thermodynamics, are ``irreversible''. Seen from the molecular-mechanical point of view, however all courses of events are reversible. If one calls a molecular-theoretically defined state a microscopically described one, or, more briefly, micro-state, then an immensely large number (Z) of states belong to a macroscopic condition. Z is then a measure of the probability of a chosen macro-state. This idea appears to be of outstanding importance also because of the fact that its usefulness is not limited to microscopic description on the basis of mechanics.''
	\end{quote}
	
	\begin{flushright}
	--- A. Einstein, Autobiographical notes
	\end{flushright}
	
	\ 
	
	\noindent\textbf{\underline{Acknowledgements}}
	
	I thank S. Goldstein and E. Speer for many very useful comments. I also thank A.J. Krueger for very helpful technical support in putting this paper together.
	
	\

	\end{document}